\documentclass[aps,pra,groupaddress,showpacs,twocolumn]{revtex4-1}
\usepackage{amsfonts}
\usepackage{bm}
\usepackage{epsfig,amsmath,graphicx,amssymb,overpic}
\usepackage{color}

\def\be{\begin{equation}}
\def\ee{\end{equation}}
\def\bee{\begin{eqnarray}}
\def\ene{\end{eqnarray}}
\def\bes{\begin{subequations}}
\def\ees{\end{subequations}}

\newcommand{\bv}{{\bf v}}

\newcommand{\PT}{{\cal PT}}

\begin{document}

\baselineskip=12pt

\title{Spatial solitons and stability in self-focusing and defocusing Kerr nonlinear media \\ with generalized $\PT$-symmetric Scarff-II potentials}
\author{Zhenya Yan$^{1,2}$}
\email{zyyan@mmrc.iss.ac.cn}
\author{Zichao Wen$^{1}$}
\author{Chao Hang$^{3,4}$}
\affiliation{$^{1}$Key Laboratory of Mathematics Mechanization, Institute of Systems
Science, AMSS, Chinese Academy of Sciences, Beijing 100190, China \\
$^{2}$State Key Laboratory of Theoretical Physics, Institute of Theoretical Physics, Chinese Academy of Sciences, Beijing 100190, China \\
$^{3}$State Key Laboratory of Precision Spectroscopy and Department of Physics,
                 East China Normal University, Shanghai 200062, China\\
$^4$NYU-ECNU Institute of Physics at NYU Shanghai, Shanghai, 200062, China}

\begin{abstract}
We present a unified theoretical study of the bright solitons governed by self-focusing and defocusing nonlinear Schr\"odinger (NLS) equations with generalized parity-time ($\PT$)-symmetric Scarff II potentials. Particularly, a $\PT$-symmetric $k$-wavenumber Scarff II potential and a multi-well Scarff II potential are considered, respectively. For the $k$-wavenumber Scarff II potential, the parameter space can be divided into different regions, corresponding to unbroken and broken $\PT$-symmetry and the bright solitons for self-focusing and defocusing Kerr nonlinearities. For the multi-well Scarff II potential the bright solitons can be obtained by using a periodically space-modulated Kerr nonlinearity. The linear stability of bright solitons with $\PT$-symmetric $k$-wavenumber and multi-well Scarff II potentials is analyzed in details  using numerical simulations. Stable and unstable bright solitons are found in both regions of unbroken and broken $\PT$-symmetry due to the existence of the nonlinearity. Furthermore, the bright solitons in three-dimensional self-focusing and defocusing NLS equations with a generalized $\PT$-symmetric Scarff II potential are explored. This may have potential applications in the field of optical information transmission and processing based on optical solitons in nonlinear dissipative but $\PT$-symmetric systems.

\end{abstract}

\maketitle

\section{Introduction}

The nonlinear Schr\"odinger (NLS) equation plays an important role in many fields of nonlinear physics~\cite{nls1,nls2,nls3,nls4,nls5}. The cubic NLS equation is shown to be completely integrable for both self-focusing and defocusing Kerr nonlinearities~\cite{c1,c2,c3}, which can be used to describe the propagation of optical pulses in Kerr-type optical media~\cite{nls3,nls4,nls5}, or to describe the dynamics of matter waves in Bose-Einstein condensates, known as the Gross-Pitaevskii equation~\cite{bec1,bec2,bec3,bec4}.

The NLS equations with real external potentials and gain-and-loss distributions  have been studied in many works~\cite{simi0, yanpla10, yanpre12,1dnlsv,1dnlsv2} because the refractive index of optical materials is usually complex, i.e., $n(x)=n_R(x)+in_I(x)$ with $n_R(x)$ and $n_I(x)$ being the real and imaginary  parts, respectively. In optics, the propagation of a signal is stable unless the propagation constant of the light is in real spectrum range. This requirement can be efficiently achieved if the gain-and-loss distributions in the medium are exactly balanced to ensure the relation $n(x)=n^{\ast}(-x)$ (or $n_R(x)=n_R(-x)$ and $n_I(-x)=-n_I(x)$), which is known as the parity-time ($\PT$) symmetric  systems~\cite{bender}. $\PT$-symmetry may exhibit entirely real spectrum of the respective optical potential in some parameter regions, referred to as the unbroken $\PT$-symmetry~\cite{bender,sh2,sh1}. Beyond these regions, in the broken $\PT$-symmetry, the spectrum becomes complex and propagating waves may be either grow or decay.

Recently, various $\PT$-symmetric potentials have been introduced to the NLS equations, which have been shown to possess stable and unstable solitons of different types~\cite{ptn, ptn0, ptn2,dnls,per1,per2,morse,harm1,harm2,gau,gau2,pt1,pt2,pt3,pt4,pt5,pt6,pt7}. Examples include the NLS equations with $\PT$-symmetric Scarff II potential~\cite{ds,ptn,ptn0,ptn2,dnls}, periodic potential~\cite{ptn,ptn0,per1,per2}, harmonic potential~\cite{harm1,harm2}, Rosen-Morse potential~\cite{morse}, Gaussian potential~\cite{harm1,gau,gau2}, sextic anharmonic double-well potential~\cite{anharm}, time-dependent harmonic-Gaussian potential~\cite{yanpra15}, self-induced potential~\cite{yanaml}, and etc. (see, e.g., \cite{pt1,pt2,pt3,pt4,pt5,pt6,pt7}). One the other hand, due to the significant progress achieved in recent years on developing optical materials with adjustable refractive index, $\mathcal{PT}$-symmetric optical systems made of solid-state waveguides and fiber networks~\cite{Guo,Feng,Regens}, multi-level atomic systems~\cite{HHK,Sheng,HZKH,HZKHM,Hang}, and microcavities~\cite{Peng,Chang} have been suggested or realized experimentally. These practical systems provide a solid ground for the study of solitons in the NLS model with $\PT$-symmetric potentials.

The $\PT$-symmetric potentials play a key role in the wave propagation in the NLS equations in fibre and waveguide optics. As an example, the wave propagation with a $\PT$-symmetric Scarff II potential in both linear and nonlinear models has been studied in recent years~\cite{ptn,ptn0,ptn2,dnls}. The $\PT$-symmetric Scarff II potential can support stable bright solitons in the NLS models within particular parameter regions~\cite{sh1,sh2}. More importantly, the real and imaginary parts of $\PT$-symmetric $k$-wavenumber Scarff II potential~\cite{ksc,ksc2} both approach to zero as $|x|\to \infty$ [cf. Eq.~(\ref{poten}), for $V_0=2,\, W_0=1,\, k=\sqrt{2}$, we have $V(x)\approx 3.25\times 10^{-24}$ and $W(x)\approx 1.04\times 10^{-12}$ as $x=20$], that is to say, they have limited effect on the nonlinear waves for the NLS equation only in a boundary region (e.g., $|x|<20$ for $k=\sqrt{2}$) [cf. Eq.~(\ref{ode})] and can more easily support the existence of bright solitons in the NLS equation, compared with the $\PT$-symmetric harmonic and optical lattice potentials, which always have the effect on the nonlinear waves in the whole region  (see, e.g, \cite{ptn,ptn0,per1,per2, harm1,harm2}). However, a complete analysis of the soliton stability in a $\PT$-symmetric $k$-wavenumber Scarff II potential is still lacking in a full parameter space (e.g., the space consists of amplitudes of real and imaginary parts of $\PT$-symmetric $k$-wavenumber Scarff II potential and the wavenumber $k$, see Eq.~(\ref{poten})). In addition, the study of optical solitons could be very useful in the information science. For example, one can use optical solitons in coding for secured optical communication, etc.

In this work we propose a unified study of the optical bright solitons governed by self-focusing and defocusing nonlinear Schr\"odinger (NLS) equations with generalized $\PT$-symmetric Scarff II potentials. Particularly, we consider a $\PT$-symmetric $k$-wavenumber Scarff II potential and a multi-well Scarff II potential, respectively. For the $k$-wavenumber Scarff II potential, we show that by using two $k$-wavenumber rays and a $k$-wavenumber parabola one can divide the parameter space $\{(V_0, W_0)| V_0>0, W_0\in\mathbb{R}\}$ ($V_0$ and $W_0$ are, respectively, the real and imaginary parts of the potential) in different regions, corresponding to unbroken and broken $\PT$-symmetry and the bright solitons with self-focusing and defocusing Kerr nonlinearities. For the multi-well Scarff II potential we obtain bright solitons by using a periodically space-modulated Kerr nonlinearity. Then, we analyze the linear stability of bright solitons for both potentials in details by using numerical simulations. Surprisingly, stable and unstable bright solitons are found in both regions of unbroken and broken $\PT$-symmetry due to the existence of the nonlinearity. Furthermore, we explore the bright solitons in three-dimensional (3D) self-focusing and defocusing NLS equations with a generalized $\PT$-symmetric Scarff II potential.

Turning to the possible experimental implementation of
the models we will study [see Eq. (\ref{nls}) with $\PT$-symmetric $k$-wavenumber and multi-well Scarff II potentials (\ref{poten}) and (\ref{poten2}) ], we can exploit a $\PT$-symmetric refractive index profile (e.g., the considered $\PT$-symmetric $k$-wavenumber Scarff II potential) imprinted in a
cold gas of two atomic isotopes of $\Lambda$-type configuration (say of $^{87}$Rb and $^{85}$Rb isotopes) loaded in an atomic cell, as suggested in Refs.~\cite{HHK,Sheng,HZKH,HZKHM,Hang}. By the interference of two
Raman resonances, the required spatial distribution of the refractive index (i.e., the $k$-wavenumber and multi-well Scarff II potentials) can be achieved by a proper combination of
a control laser field and a far-off-resonance Stark
laser field. Since the refractive index of the atomic vapor is determined by two external laser fields, whose intensities can be lowered to several microwatts achievable for the lasers nowadays, the system has the advantages of real-time all-optical tunable capability and controllable $\PT$-symmetry accuracy. In addition, the studied model supports a large Kerr nonlinearity (which is at least $10^{13}$-order larger than those measured for usual nonlinear optical materials) due to the Raman resonant character, which favors to the formation of optical bright solitons.

The rest of this paper is arranged as follows. Sec.~II gives the bright spatial solitons in one-dimensional $\PT$-symmetric potentials. Both $\PT$-symmetric $k$-wavenumber and multi-well Scarff II potentials are considered. The linear stability of optical bright solitons is analyzed. Stable solitons are found for both unbroken and broken $\PT$-symmetric potentials. Sec.~III gives the bright solitons in 3D self-focusing and defocusing NLS equations with a generalized $\PT$-symmetric Scarff II potential. Finally, the last section contains a summary of main results obtained in this work.

\section{Bright solitons in one-dimensional $\PT$-symmetric potentials}

\subsection{Nonlinear physical model and stationary solutions}

Now we uniformly investigate the exact localized solution modes and their stabilities of self-focusing and defocusing NLS equations with a $\PT$-symmetric potential. In the dimensionless form, the physical model can be written as~\cite{ptn}
\begin{eqnarray}
\label{nls}
 i\partial_t\psi+\partial_x^2\psi+[V(x)+iW(x)]\psi+g|\psi|^2\psi=0,
 \end{eqnarray}
where $\partial_t=\partial/\partial t,\, \partial_x=\partial/\partial x$,\, $\psi\equiv\psi(x,t)$ is a complex field of $x, t$ ($x$ and $t$ are, respectively, dimensionless space and time), $V(x)$ is a real external potential, $W(x)$ is a real gain-and-loss distribution,
$g$ characterizes the self-focusing ($g>0)$ or defocusing $(g<0)$ Kerr nonlinearity, respectively.   Eq.~(\ref{nls}) is associated with a variational principle $\delta\mathcal{L}(\psi)/\delta\Psi^{*}=0$ with the Lagrangian density
\bee
\nonumber\mathcal{L}(\psi)=i\left(\psi^{*}\psi_t-\psi\psi^{*}_t\right) +2|\psi_x|^2 \qquad\qquad \\
+2[V(x)+iW(x)]|\psi|^2+g|\psi|^4.
\ene
The $\PT$-symmetric potential $V(x)+iW(x)$ leads to the sufficient (but not necessary) condition $V(x)=V(-x)$ and $W(x)=-W(-x)$. The quasipower and power of Eq.~(\ref{nls}) are given by $Q(t)=\int_{-\infty}^{+\infty}\psi(x,t)\psi^{*}(-x,t)dx$ and $P(t)=\int_{-\infty}^{+\infty}|\psi(x,t)|^2ds$, respectively. One can readily get that $Q_t=i\int_{-\infty}^{+\infty}g\psi(x,t)\psi^{*}(-x,t)[|\psi(x,t)|^2-|\psi(-x,t)|^2]dx$ and $P_t=
-2\int_{-\infty}^{+\infty}W(x)|\psi(x,t)|^2dx$. Notice that if we set $t\to z$ in Eq.~(\ref{nls}), where $z$ denotes the propagation distance~\cite{ptn}, then the following results about Eq.~(\ref{nls}) still hold as $t\to z$.

We focus on the stationary solutions of Eq.~(\ref{nls}) in the form $\psi(x, t)=\phi(x)e^{i\mu t}$, where $\mu$ is the real propagation constant and the complex function $\phi(x)$ satisfies the stationary NLS equation with varying parameter modulation
\bee
\mu\phi(x)\!=\!\frac{d^2\phi(x)}{dx^2}\!+\![V(x)\!+\! iW(x)]\phi(x)\!+\!g|\phi(x)|^2\phi(x),\quad
\label{ode}
\ene
which can be solved by using numerical methods.

In the following, we consider two different types of $\PT$-symmetric potential, i.e., the $\PT$-symmetric $k$-wavenumber Scarff II potential and the periodically space-modulated Scarff II potential, and study the exact localized solutions and their linear stability in both regions of unbroken and broken $\PT$-symmetry.

\subsection{$\PT$-symmetric $k$-wavenumber Scarff II potential}

\begin{figure}[!t]
\begin{center}
\hspace{-0.1in}{\scalebox{0.42}[0.45]{\includegraphics{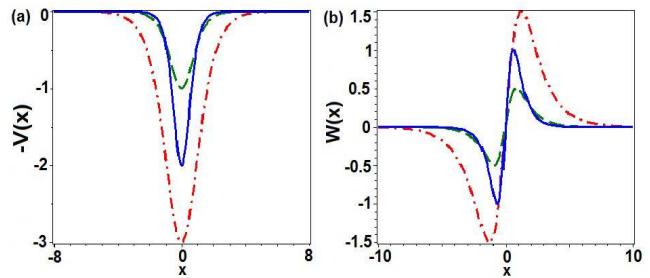}}}
\end{center}
\vspace{-0.15in} \caption{\small (color online) Real (a) and imaginary (b) parts of $\PT$-symmetric $k$-wavenumber Scarff II potential (\ref{poten}). (a): $V_0=3,\, k=1/\sqrt{2}$ (dashed-dotted line),  $V_0=2,\, k=\sqrt{2}$ (solid line),  $V_0=k=1$ (dashed line), and (b): $W_0=3,\, k=1/\sqrt{2}$ (dashed-dotted line),  $W_0=2,\, k=\sqrt{2}$ (solid line),  $W_0=k=1$ (dashed line).} \label{fig-1g}
\end{figure}

To study the soliton solutions of Eq.~(\ref{ode}), we first consider the $\PT$-symmetric $k$-wavenumber Scarff II potential~\cite{ksc,ksc2}, which is given as
\bee \label{poten}
 V(x)\!=\!V_0\,{\rm sech}^2\!(kx), \,\, W(x)\!=\!W_0\,{\rm sech}\!(kx)\tanh(kx),\,\,\,
\ene
where $k>0$ denotes the wavenumber, and $V_0>0$ and $W_0$ are real parameters and modulate amplitudes of the external potential and
gain-and-loss distribution, respectively. The wavenumber $k$ and amplitude $V_0$ can modulated the well width and depth of the potential $V(x)$, respectively. $W_0$ can modulate the impact of the gain-and-loss distribution. For the case $k=1$, $\PT$-symmetric $k$-wavenumber Scarff II potential (\ref{poten}) becomes the usual $\PT$-symmetric Scarff II potential~\cite{sh2}. Figure~\ref{fig-1g} displays the $\PT$-symmetric $k$-wavenumber Scarff II potential (\ref{poten}) for different amplitudes $V_0,\, W_0$ and wavenumber $k$.

It follows from Eq.~(\ref{poten}) that $V(x),\, W(x)\to 0$ as $|x|\to \infty$ and/or $k\to \infty$, and $V(x)\to V_0,\, W(x)\to 0$ as $x=0$ and/or
$k\to 0$. If $V_0W_0\not=0$, then $V(x)$ and $W(x)$ satisfy the family of elliptic curves: \bee
\left(\frac{V(x)}{V_0}-\frac12\right)^2+\frac{W^2(x)}{W_0^2}=\frac14
\ene
with the centers being $(V_0/2, 0)$ in the $(V(x), W(x))$-space . It is easy to see that the family of elliptic curves is dependent on the wavenumber $k$.

In general, for the $\PT$-symmetric $k$-wavenumber Scarff II potential $S_{\PT}=V(x)+iW(x)$ with  $V(x)$ and
$ W(x)$  given by Eq.~(\ref{poten}) for any real parameters $V_0, W_0$ there are the following three cases:
\begin{itemize}
\item[i)] When $W_0=0$, the potential $S_{\PT}=V(x)$ is a real, Hermitian, and $\PT$-symmetric potential;

\item[ii)] When $V_0=0$, the potential $S_{\PT}=iW(x)$ is a pure imaginary, non-Hermitian, and $\PT$-symmetric potential;

\item[iii)] When $V_0W_0\not=0$, the potential $S_{\PT}=V(x)+iW(x)$ is a complex, non-Hermitian, and $\PT$-symmetric potential.
\end{itemize}

In the following, we mainly consider the potential $S_{\PT}=V(x)+iW(x)$ in Case iii) with $V_0>0$.

\subsubsection{linear eigenvalue problem with unbrokon/brokon $\PT$-symmetry}

\begin{figure}
\begin{center}
\hspace{-0.1in}{\scalebox{0.42}[0.4]{\includegraphics{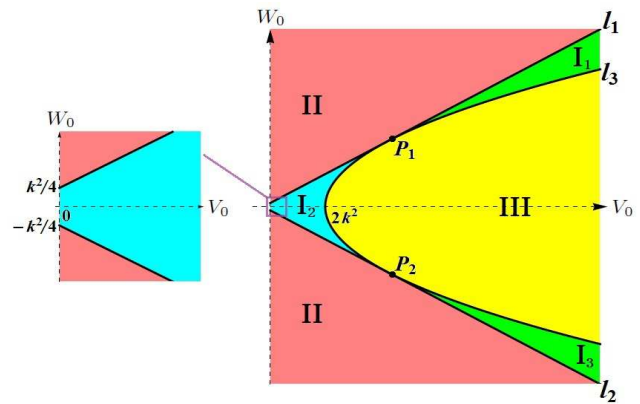}}}
\end{center}
\vspace{-0.15in} \caption{\small (color online) The parameter space $\{(V_0, W_0)| V_0>0,\, W_0\in \mathbb{R}\}$ can be divided in different regions by two families of $k$-wavenumber open rays $l_{1,2}:\, W_0=\pm(V_0+k^2/4)$ with $V_0>0$ and a family of $k$-wavenumber parabola $l_3:\, W_0^2=9k^2(V_0-2k^2)$, corresponding to unbroken and broken $\PT$-symmetry and bright solitons with self-focusing ($g>0)$ and defocusing ($g<0$) Kerr nonlinearities. The tangent points of three curves are $P_{1,2}=(17k^2/4, \pm 9k^2/2)$. } \label{fig-1}
\end{figure}

\begin{table*}[!t]
\vspace{-0.1in}
\caption{\small The regions for unbroken and broken $\PT$-symmetry and the existence of bright solitons with self-focusing ($g>0)$ and defocusing ($g<0$) nonlinearities in the parameter space $\{(V_0, W_0)| V_0>0, W_0\in \mathbb{R}\}$ [The signs `Yes' and `No' denote the corresponding problems exist and do not exist in the regions, respectively. The signs  `+' and `-' stand for the union and difference of two sets, respectively]. \vspace{0.05in}}
\begin{tabular}{ccccc} \hline\hline \\ [-2.0ex]
\noindent  Region  &   Linear problem ($g=0$):  &  Linear problem ($g=0$):
& Nonlinear problem ($g>0$):  &  Nonlinear problem ($g<0$):  \\  [0.5ex]
\noindent   & \quad   unbroken $\PT$-symmetry \quad  & broken $\PT$-symmetry  &  soliton (\ref{solu})   &  soliton (\ref{solu})   \\ [1.0ex] \hline \\ [-2.0ex]
 $\begin{array}{l}
  {\rm I}_1+{\rm I}_2+{\rm I}_3+l_1\\
   +l_2 -\{P_{1,2}\}\end{array}$ &   Yes   & No & Yes & No \\  [2.0ex] \hline \\ [-1.6ex]
  $l_3$ &   Yes  & No & No & No  \\  [1.0ex] \hline \\ [-2.0ex]
  II &   No  & Yes & Yes & No  \\  [1.0ex] \hline \\ [-2.0ex]
  III &   Yes  & No & No & Yes  \\ [1.0ex] \hline\hline
\end{tabular}
\label{table}
\end{table*}

\begin{figure*}[!t]
\begin{center}
\hspace{-0.1in}{\scalebox{0.75}[0.75]{\includegraphics{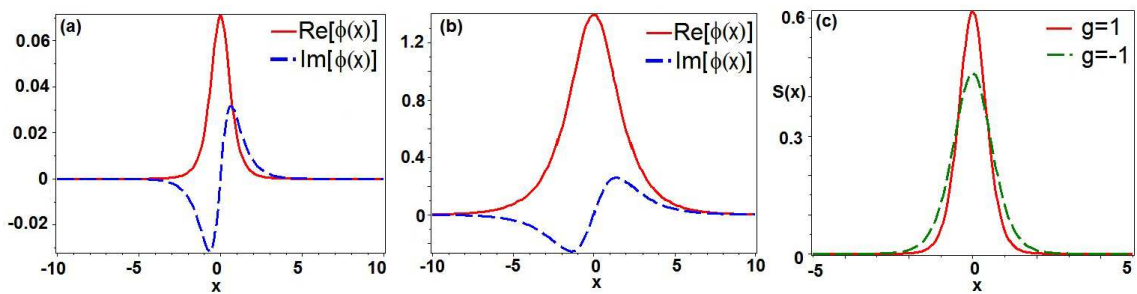}}}
\end{center}
\vspace{-0.15in} \caption{\small (color online) Real (solid lines) and imaginary (dashed lines) parts of bright soliton (\ref{solu}) for (a) the self-focusing nonlinearity $g=1$ with $V_0=5,\, W_0=5.2,\ k=\sqrt{2}$, and (b) defocusing nonlinearity $g=-1$ with $V_0=3,\, W_0=0.5,\, k=1/\sqrt{2}$. (c) The transverse power-flow vector (Poynting vector) $S(x)$ given by Eq.~(\ref{s}) for the self-focusing nonlinearity $g=1$ with $V_0=5,\, W_0=5.2,\ k=\sqrt{2}$ (solid line) and defocusing nonlinearity $g=-1$ with $V_0=3,\, W_0=0.5,\, k=1/\sqrt{2}$ (dashed line).} \label{fig-2}
\end{figure*}

The linear eigenvalue problem
\bee
L_{\PT}\Phi(x)=\lambda\Phi(x), \,\, L_{\PT}=-\partial_x^2-[V(x)+iW(x)]\,\,
 \ene
with $V(x)$ and $W(x)$ given by Eq.~(\ref{poten}) exhibits an entirely real spectrum provided that two amplitudes $V_0>0,\, W_0$ and the wavenumber $k$ satisfy
\bee \label{ptc}
|W_0|\leq V_0+\frac{k^2}{4},
\ene
where $\lambda$ and $\Phi(x)$ are eigenvalue and engenfunction, respectively. The inequality (\ref{ptc}) is called the $\PT$ symmetry-unbreaking condition. If $|W_0|>V_0+k^2/4$, the spectrum becomes complex. For $k=1$, the condition (\ref{ptc}) reduces to the well-known one, $|W_0|\leq V_0+1/4$~\cite{sh1,sh2}. Therefore, two families of $k$-wavenumber open rays $l_1:$
 \bee \label{l1}
  W_0=V_0+k^2/4,\quad V_0>0
  \ene
 and $l_2:$
 \bee\label{l2}
  W_0=-(V_0+k^2/4), \quad V_0>0
  \ene
 divide the parameter space $\{(V_0, W_0)| V_0>0, W_0\in\mathbb{R}\}$ into two regions, i.e., the unbroken $\PT$-symmetric region $|W_0|\leq V_0+k^2/4$ and the broken $\PT$-symmetric region $|W_0|> V_0+k^2/4$ (see Fig.~\ref{fig-1}).

\subsubsection{The conditions for the existence of nonlinear modes}

For the given $\PT$-symmetric $k$-wavenumber Scarff II potential (\ref{poten}) with $V_0>0$ and $W_0\in \mathbb{R}$, Eq.~(\ref{ode}) admits  the unified bright soliton for both self-focusing ($g>0)$ and defocusing $(g<0)$ nonlinearities
\bee\label{solu}
\phi(x)=\sqrt{\frac{1}{g}\left(\frac{W_0^2}{9k^2}-V_0+2k^2\right)}\,{\rm sech}(kx)e^{i\varphi(x)},
\ene
where $\mu=k^2$,\, $g[W_0^2/(9k^2)-V_0+2k^2]>0$, and the phase is
\bee
\varphi(x)=\frac{W_0}{3k^2}\arctan[\sinh(kx)].
 \ene
 In particular, when $g=k=1$ or $-g=k=1$, the solution (\ref{solu}) reduces to the known ones as given in Refs.~\cite{ptn, dnls}.

Now we analyze the conditions for the existence of bright soliton (\ref{solu}) for parameters $W_0,\, V_0,\, k$, and $g$. For the positive (self-focusing) nonlinearity $g>0$, the condition for the existence of bright soliton (\ref{solu}) is given by
\be \label{con1}
W_0^2>9k^2(V_0-2k^2).
\ee
The real and imaginary parts of the soliton are displayed in Fig.~\ref{fig-2}(a). For the negative (defocusing) nonlinearity $g<0$, the condition for the existence of bright soliton (\ref{solu}) is given by
\be \label{con2}
|W_0|<3k\sqrt{V_0-2k^2}, \,\,\,\, V_0>2k^2.
\ee
The real and imaginary parts of the soliton are illustrated in Fig.~\ref{fig-2}(b).

Taking into account the conditions (\ref{con1}) and (\ref{con2}) for the existence of bright soliton (\ref{solu}) for $g>0$ and $g<0$, the parameter space $\{(V_0, W_0)| V_0>0, W_0\in\mathbb{R}\}$ can be divided into another two fundamental regions by the family of parabolas $l_3:\, W_0^2=9k^2(V_0-2k^2)$ with vertex being $(2k^2, 0)$. These two conditions (\ref{con1}) and (\ref{con2}) for bright solitons with $g>0$ and $g<0$, respectively, are shown in Fig.~\ref{fig-1}.

Notice that the bright soliton (\ref{solu}) with $g>0$ (attractive case for Eq.~(\ref{nls})) can also exist for the region $\{(V_0, W_0)| V_0\leq 0, \, W_0\in\mathbb{R}\}$ even if we do not study the $\PT$-unbroken/broken of the linear eigenvalue problem in this region.

Two branches of $k$-wavenumber open rays $l_{1,2}$ given by Eqs.~(\ref{l1}) snd (\ref{l2}) are tangent to the parabola $l_3:$
\bee
 W_0^2=9k^2(V_0-2k^2)
 \ene
 at the points $P_{1,2}=(17k^2/4, \pm 9k^2/2)$, respectively. In another word, $|V_0+k^2/4|>3k\sqrt{V_0-2k^2}$ always holds except for two tangent points $P_{1,2}$, at which they are equal (see Fig.~\ref{fig-1} and note~\cite{tangent} for the details).

Therefore, the parameter space $\{(V_0, W_0)| V_0>0,\, W_0\in\mathbb{R}\}$ can be divided into eight regions, i.e., \bee\begin{array}{l}
 \{(V_0, W_0)| V_0>0, \, W_0\in\mathbb{R}\} \vspace{0.1in} \\
  \qquad ={\rm I}_1+{\rm I}_2+{\rm I}_3+{\rm II}+{\rm III}+l_1+l_2+l_3,
  \end{array}
  \ene
  where `$+$' denotes the union of two sets, these regions ${\rm I}_{1,2,3},\,{\rm II},\, {\rm III}$ denote open sets without the corresponding boundaries (see Fig.~\ref{fig-1}). To make it become clearer, we summarize these different regions in Table~\ref{table}. One can see that the region for bright solitons with self-focusing nonlinearity contains both regions for unbroken and broken $\PT$-symmetry, whereas the region for bright solitons with defocusing nonlinearity contains only the region for unbroken $\PT$-symmetry.

For any complex solution $\phi(x)=|\phi(x)|e^{i\varphi(x)}$ with $\varphi(x)$ bering a real phase of $\phi(x)$, we find that its corresponding transverse power-flow or Poynting vector and the gradient of phase obey the relation $S(x)=\frac{i}{2}(\phi\phi_x^{*}-\phi^{*}\phi_x)=|\phi(x)|^2\varphi_x$. For the bright soliton (\ref{solu}), its corresponding transverse power-flow or `Poynting vector' is given by \bee
\label{s}
S(x)=\frac{W_0}{3kg}\left(\frac{W_0^2}{9k^2}-V_0+2k^2\right){\rm sech}^3(kx),
 \ene
 which implies that ${\rm sgn}(S(x))={\rm sgn}(W_0)$ for any $x$ and $g=\pm 1$ (see Fig.~\ref{fig-2}(c)) since
 $k>0$ and $g(W_0^2/(9k^2)-V_0+2k^2)>0$ are required for the conditions of the existence of bright solitons. Therefore, the power always flows in one direction, i.e., from the gain toward the loss. In addition, we have the conserved power related to solution (\ref{solu}) is  $P(t)=\frac{2}{gk}\left(\frac{W_0^2}{9k^2}-V_0+2k^2\right)$.

\begin{figure}[!t]
\begin{center}
\vspace{0.05in}	
\hspace{-0.1in}{\scalebox{0.58}[0.58]{\includegraphics{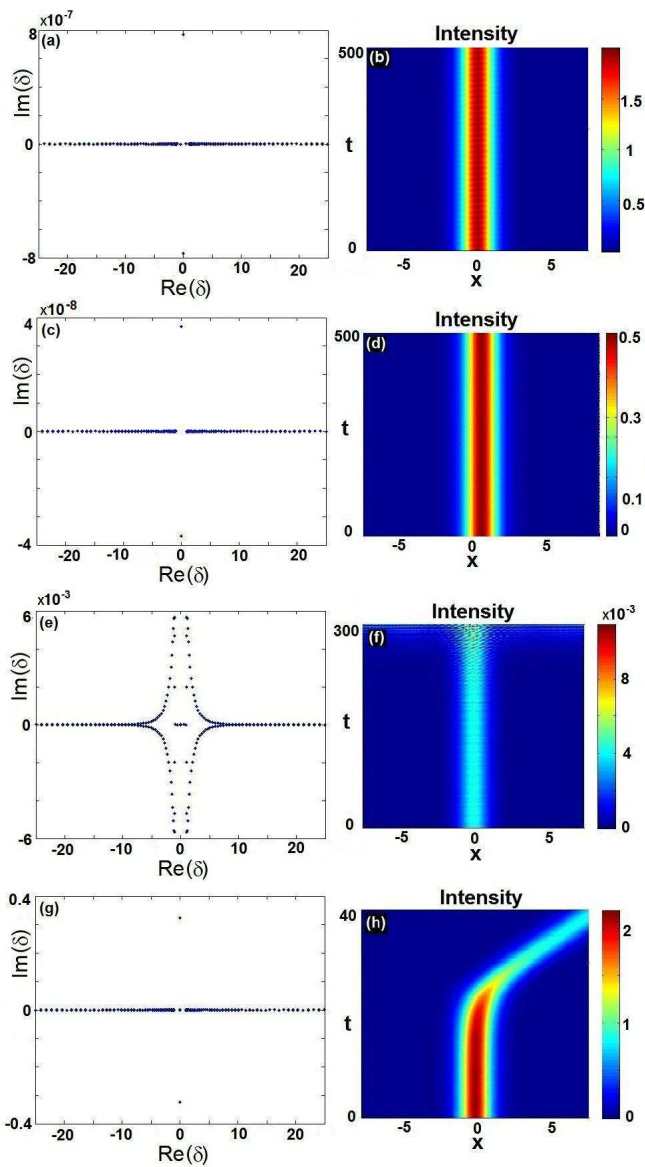}}}
\end{center}
\vspace{-0.15in} \caption{\small (color online)  The linear stability eigenvalues (left column) and propagation of soliton intensity $|\psi(x,t)|^2$ with $g=k=1$ (right column). (a)-(b) $V_0=0.1,\, W_0=0.5$ (in the region II), (c)-(d) $V_0=1.5,\, W_0=0.2$ (in the region ${\rm I}_2$), (e)-(f) $V_0=5,\, W_0=5.2$ (in the region ${\rm I}_1$), and (g)-(h) $V_0=-0.1,\, W_0=0.1$ (in the region $\{(V_0, W_0)| V_0\leq 0, \, W_0\in\mathbb{R}\}$).} \label{fig3}
\end{figure}

\subsubsection{Stability of nonlinear modes}

Next we study the linear stability of the bright solitons (\ref{solu}) in the above-mentioned different regions. To this end, we considered a perturbed bright soliton solution~\cite{stable}
\begin{equation}
\label{pert}
\psi(x,t)=\left\{\phi(x)+\epsilon \left[F(x)e^{i\delta t}\!+\!G^*(x)e^{-i\delta^* t}\right]\right\}e^{i\mu t},
\end{equation}
where $\phi(x)e^{i\mu t}$ is a stationary solution of Eq.~(\ref{nls}), $\epsilon\ll 1$, and $F(x)$ and $G(x)$ are the eigenfunctions of the linearized eigenvalue problem. Substituting Eq.~(\ref{pert}) in Eq.~(\ref{nls}) and linearizing with respect to $\epsilon$, we obtain the following linear eigenvalue problem
\begin{eqnarray}
\left(\begin{array}{cc}   L & g\phi^2(x) \\   -g\phi^{*2}(x) & -L^* \\  \end{array}\right)
\left(  \begin{array}{c}    F(x) \\    G(x) \\  \end{array} \right)
=\delta \left(  \begin{array}{c}   F(x)\\    G(x) \\  \end{array}\right),
\label{stable}
\end{eqnarray}
where $L=\partial^2_x+V(x)+iW(x)+2g|\phi(x)|^2-\mu$.

The stability of the perturbed soliton $\psi(x,t)$ is related to the imaginary parts ${\rm Im}(\delta)$ of all eigenvalues $\delta$. If $|{\rm Im}(\delta)|>0$, then the solution $\psi(x,t)$ will grow exponentially with $t$ (i.e., it is unstable), otherwise the solution $\psi(x,t)$ is stable.
In the following, we study the stability of exact soliton (\ref{solu}) under an initial random noise
perturbation up to $2\%$ of its amplitude for different parameters

Since the regions I$_1$ and I$_3$ are symmetric about the $V_0$-axes, we only need to consider one of them (say I$_1$). Thus we leave four regions for studying the linear stability of the bright solitons (\ref{solu}). The regions I$_1$, I$_2$ and II correspond to bright solitons for the self-focusing nonlinearity $g=1$, whereas the region III corresponds to bright solitons for the defocusing nonlinearity $g=-1$. To be specific,
we study the stability of bright solitons in the regions for unbroken and broken $\PT$-symmetry with $k=1$ (the usual $\PT$-symmetric Scarff II potential) and $k\not=1$. Notice that we only consider $W_0>0$ as $W(x)$ is an odd function.

\begin{figure}[!t]
\begin{center}
\hspace{-0.1in}{\scalebox{0.43}[0.43]{\includegraphics{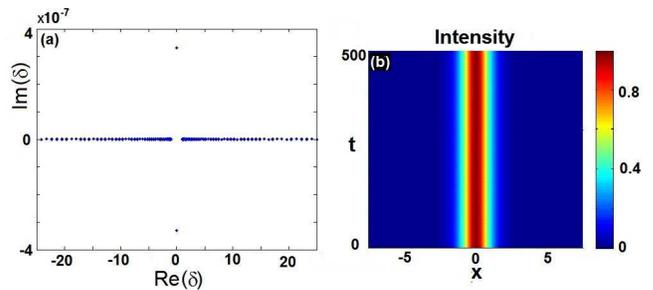}}}
\end{center}
\vspace{-0.15in} \caption{\small (color online) (a) The linear stability eigenvalues; (b) The stable propagation of soliton intensity $|\psi(x,t)|^2$ with $g=-1$. The other parameters are $V_0=3,\, W_0=0.5,\, k=1$ (in the region III).} \label{fig3c}
\end{figure}

If $k=1$ and $g=1$: (i) for $V_0=0.1,\, W_0=0.5$ in the region II (broken $\PT$-symmetry), we numerically obtain  the spectrum of eigenvalues $\delta$ (see Fig.~\ref{fig3}(a)) and a stable propagation of the soliton intensity $|\psi(x,t)|^2$ shown in Fig.~\ref{fig3}(b);
(ii) For $V_0=1.5,\, W_0=0.2$ in the region ${\rm I}_2$ (unbroken $\PT$-symmetry), we numerically obtain the spectrum of eigenvalues $\delta$ (see Fig.~\ref{fig3}(c)) and a stable propagation of the soliton intensity $|\psi(x,t)|^2$
shown in Fig.~\ref{fig3}(d); (iii) For $V_0=5,\, W_0=5.2$ in the region ${\rm I}_1$ (unbroken $\PT$-symmetry), we numerically obtain the spectrum of eigenvalues $\delta$ (see Fig.~\ref{fig3}(e)) and an unstable propagation of the soliton intensity $|\psi(x,t)|^2$  shown in Fig.~\ref{fig3}(f); (iv) For $V_0=-0.1,\, W_0=0.1$ in the region $\{(V_0, W_0)| V_0\leq 0, \, W_0\in\mathbb{R}\}$, we numerically obtain  the spectrum of eigenvalues $\delta$ (see Fig.~\ref{fig3}(g)) and an unstable propagation of the soliton intensity $|\psi(x,t)|^2$ shown in Fig.~\ref{fig3}(h). If $k=1$ and $g=-1$, for $V_0=3,\, W_0=0.5$ in the region III (unbroken $\PT$-symmetry), we numerically obtain the spectrum of eigenvalues $\delta$ (see Fig.~\ref{fig3c}(a)) and a stable propagation of the soliton intensity $|\psi(x,t)|^2$ shown in Fig.~\ref{fig3c}(b). We stress that the stability (instability) of the bright solitons in broken (unbroken) $\PT$-symmetry region is due to the existence of Kerr nonlinearity.

\begin{figure}[!t]
	\begin{center}
\hspace{-0.1in}{\scalebox{0.55}[0.55]{\includegraphics{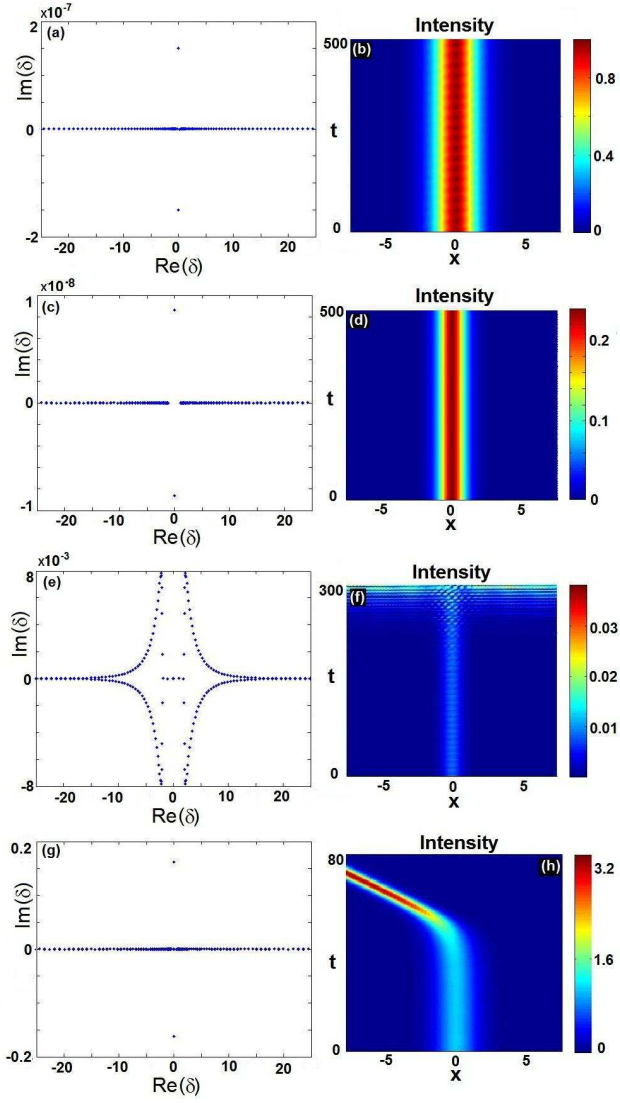}}}
	\end{center}
	\vspace{-0.15in} \caption{\small (color online)  The linear stability eigenvalues (left column) and propagation of soliton intensity $|\psi(x,t)|^2$ with $g=1$ (right column). (a)-(b) $V_0=0.05,\, W_0=0.25,\, k=1/\sqrt{2}$ (in the region II), (c)-(d), $V_0=2.16,\, W_0=0.12,\, k=\sqrt{1.2}$ (in the region ${\rm I}_2$), (e)-(f) $V_0=10,\, W_0=10.4,\, k=\sqrt{2}$ (in the region ${\rm I}_1$), and (g)-(h) $V_0=-0.05,\, W_0=0.05,\, k=1/\sqrt{2}$ (in the region $\{(V_0, W_0)| V_0\leq 0, \, W_0\in\mathbb{R}\}$.} \label{fig3k}
\end{figure}

\begin{figure}[!t]
\begin{center}
\hspace{-0.1in}{\scalebox{0.43}[0.45]{\includegraphics{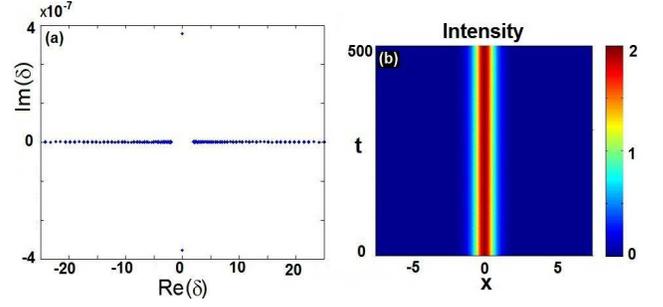}}}
\end{center}
\vspace{-0.15in} \caption{\small (color online) (a) The linear stability eigenvalues; (b) The stable propagation of soliton intensity $|\psi(x,t)|^2$ with $g=-1$. The other parameters are $V_0=6,\, W_0=1,\, k=\sqrt{2}$ (in the region III).} \label{fig3kc}
\end{figure}

If $k\not=1$ and $g=1$: (i) for $V_0=0.05,\, W_0=0.25,\, k=1/\sqrt{2}$ in the region II (broken $\PT$-symmetry), we numerically obtain the spectrum of eigenvalues $\delta$  (see Fig.~\ref{fig3k}(a)) and a stable propagation of the soliton intensity $|\psi(x,t)|^2$  shown in Fig.~\ref{fig3k}(b);
(ii) For $V_0=2.16,\, W_0=0.12,\, k=\sqrt{1.2}$ in the region ${\rm I}_2$ (unbroken $\PT$-symmetry), we numerically obtain the spectrum of eigenvalues $\delta$  (see Fig.~\ref{fig3k}(c)) and a stable propagation of the soliton intensity $|\psi(x,t)|^2$ shown in Fig.~\ref{fig3k}(d); (iii) For $V_0=10,\, W_0=10.4,\, k=\sqrt{2}$ in the region ${\rm I}_1$ (unbroken $\PT$-symmetry), we numerically obtain the spectrum of eigenvalues $\delta$  (see Fig.~\ref{fig3k}(e)) and an unstable propagation of the soliton intensity $|\psi(x,t)|^2$ shown in Fig.~\ref{fig3k}(f); (iv) For $V_0=-0.05,\, W_0=0.05,\, k=1/\sqrt{2}$, in the region $\{(V_0, W_0)| V_0\leq 0, \, W_0\in\mathbb{R}\}$, we numerically obtain the spectrum of eigenvalues $\delta$  (see Fig.~\ref{fig3k}(g)) and an unstable propagation of the soliton intensity $|\psi(x,t)|^2$  shown in Fig.~\ref{fig3k}(h). If $k\not=1$ and $g=-1$, for $V_0=6,\, W_0=1,\, k=\sqrt{2}$ in the region III (unbroken $\PT$-symmetry), we see that $|{\rm Im}(\delta)|=0$ (see Fig.~\ref{fig3kc}(a)) and a stable propagation of the soliton intensity $|\psi(x,t)|^2$ is obtained as shown in Fig.~\ref{fig3kc}(b).

\subsection{$\PT$-symmetric multi-well Scarff II potential}

\begin{figure}[!t]
\begin{center}
\hspace{-0.1in}{\scalebox{0.43}[0.45]{\includegraphics{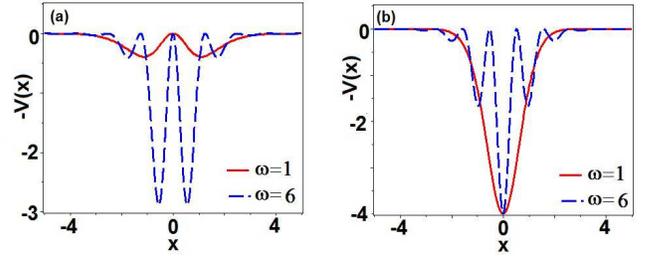}}}
\end{center}
\vspace{-0.15in} \caption{\small (color online)  The external potential $-V(x)$ exhibiting the multi-well structure for (a) $\sigma=2$ and (b) $\sigma=-2$. Other parameters are $W_0=0.2,\, \omega=1,6$.} \label{fig-cos-v}
\end{figure}

Now we consider another type of $\PT$-symmetric Scarff II potential, i.e. the $\PT$-symmetric multi-well Scarff II potential, reading
\bee\label{poten2}
\begin{array}{l}
V(x)=\left[\frac{W_0^2}{9}+2-\sigma\cos(\omega x)\right]{\rm sech}^2(x), \vspace{0.1in} \\
W(x)=W_0{\rm sech}(x)\tanh(x),
\end{array}
\ene
where $W_0,\, \sigma\in \mathbb{R}$ and $\omega\geq 0$ denotes the wavenumber. If $\omega=0$ and
$\sigma=W_0^2/9+2-V_0$, then $V(x)$ becomes the usual Scarff II potential~\cite{sh1,sh2}, which exhibits the shape of single well, but for non-zero wavenumber $\omega$, $V(x)$ exhibits the shape of multi-well, which can provide more abundant structures. For $W_0=0.2$, the external potential $V(x)$ exhibits double-well ($\sigma=2,\, \omega=1$), singe-well ($\sigma=-2,\, \omega=1$), and multi-well ($\sigma=\pm 2,\, \omega=6$) structures, respectively (see Fig.~\ref{fig-cos-v}). In fact, for the fixed parameters $W_0$ and $\omega$, one can change the parameter $\sigma$ to control the number of the wells. Notice that in the figure we plot the profile of $-V(x)$ because the Hamiltonian of Eq.~(\ref{nls}) is $-\partial_x^2-V(x)-iW(x)$.
As $\omega\to 0$ (or $\omega\ll 1$), the linear problem associated with the $\PT$-symmetric potential (\ref{poten2}) admits an entirely real spectra provided that $|W_0|\leq W_0^2/9+9/4-\sigma$. If $\omega\sim1$, the condition for the unbroken-$\PT$ symmetry is complicated. For the fixed parameter $W_0=0.2$,
we find the regions of the unbroken-$\PT$ symmetry: (i) $\omega=0.5$,\ $-280< \sigma\leq 2.5$; (ii) $\omega=1$,\, $-53.3\leq \sigma \leq 3.17$; (iii) $\omega=2$,\, $-13.1\leq\sigma\leq 6.65$; and (iv) $\omega=5$,\, $-13.9\leq\sigma\leq 11.33$.

For the $\PT$-symmetric multi-well Scarff II potential (\ref{poten2}), we consider the periodically space-modulated nonlinearity $g\to g(x)$ in Eq.~(\ref{ode}) with the form
\bee\label{g}
g(x)=\sigma\cos(\omega x).
\ene

For the zero wavenumber $\omega=0$, nonlinearity $g(x)$ is a constant, $g(x)=\sigma$, however, for the given non-zero wavenumber $\omega$, the nonlinearity $g(x)$ is either positive or negative as $x$ increases. Based on the potential (\ref{poten2}) and the nonlinearity (\ref{g}) we have  exact bright solitons of Eq.~(\ref{ode})
\bee \label{soluc}
\phi(x)={\rm sech}(x)e^{i\varphi(x)},
\ene
where $\mu=1$ and  the phase is
\bee
\varphi(x)=\frac{W_0}{3}\arctan[\sinh(x)].
 \ene
 The corresponding transverse power-flow or Poynting vector is given by
 \bee
 S(x)=\frac{i}{2}(\phi\phi_x^{*}-\phi^{*}\phi_x)=\frac{W_0}{3}{\rm sech}^3(x),
   \ene
   which implies that ${\rm sgn}(S(x))={\rm sgn}(W_0)$ for any $x$. Therefore, the power always flows in one direction, i.e., from the gain toward the loss. Moreover, the conserved power is  $P(t)=\int^{+\infty}_{-\infty}|\psi(x,t)|^2=2$.

\begin{figure}[!t]
\begin{center}
\vspace{0.1in}	
\hspace{-0.1in}{\scalebox{0.56}[0.56]{\includegraphics{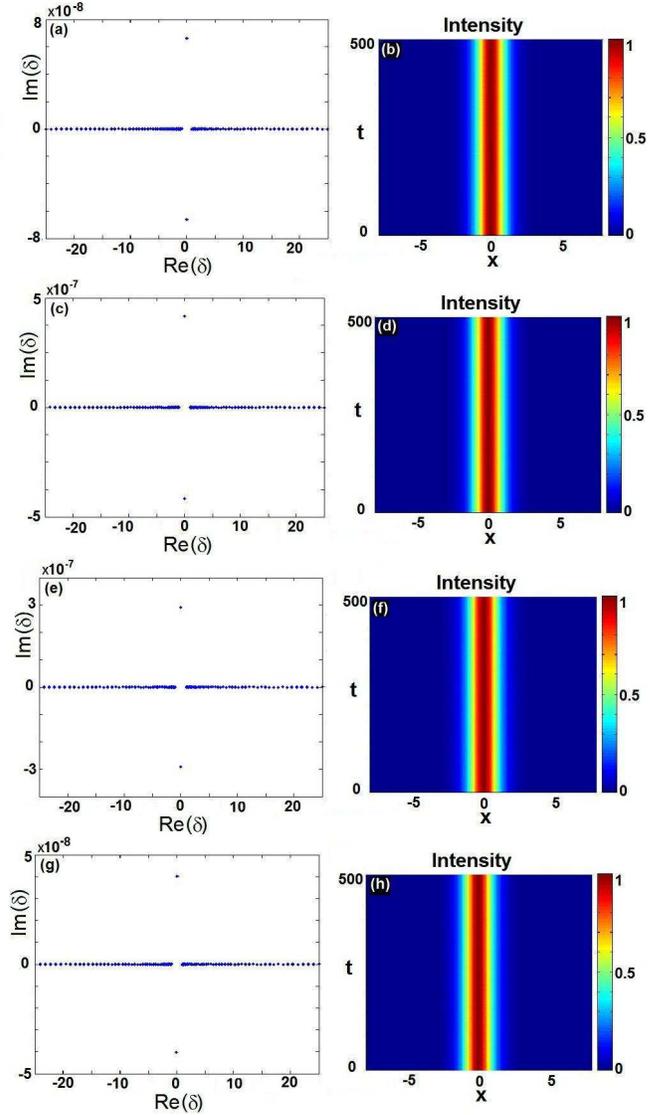}}}
\end{center}
\vspace{-0.15in} \caption{\small (color online) The linear stability eigenvalues (left column) and stable propagation of soliton intensity $|\psi(x,t)|^2$ (right column). (a)-(b) $W_0=0.2,\, \omega=1,\, \sigma=2$; (c)-(d) $W_0=0.2,\, \omega=1,\, \sigma=-2$; (e)-(f) $W_0=0.2,\, \omega=0.5,\, \sigma=-3$; (g)-(h) $W_0=0.2,\, \omega=0.5,\, \sigma=1$. All panels correspond to the unbroken $\PT$-symmetry.} \label{fig3-cos-g}
\end{figure}

Now we consider the linear stability of the soliton solution (\ref{soluc}) using Eq.~(\ref{stable}) with $g$ being replaced by the periodic function ~(\ref{g}) with $W_0=0.2$ and different values of $\omega,\, \sigma$. We find that for the fixed $\omega$ the soliton (\ref{soluc}) is more (less) stable if $\sigma<0$ ($\sigma>0$). For examples, we show the linear stability eigenvalues and the stable propagation of the soliton (\ref{soluc})
for $\omega=1,\,\sigma=2$ (see Fig.~\ref{fig3-cos-g}(a)-(b)), $\omega=1,\,\sigma=-2$ (see Fig.~\ref{fig3-cos-g}(c)-(d)), $\omega=0.5,\,\sigma=-3$ (see Fig.~\ref{fig3-cos-g}(e)-(f)), and $\omega=0.5,\,\sigma=1$ (see Fig.~\ref{fig3-cos-g}(g)-(h)).

\begin{figure}[!t]
\begin{center}
\hspace{-0.1in}{\scalebox{0.43}[0.43]{\includegraphics{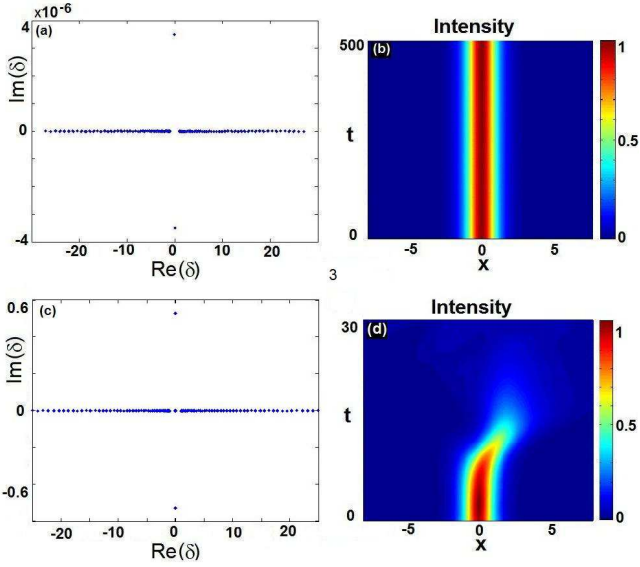}}}
\end{center}
\vspace{-0.15in} \caption{\small (color online) The linear stability eigenvalues (left column) and propagation of soliton intensity $|\psi(x,t)|^2$ (\ref{soluc}) (right column). (a)-(b) $\sigma=-310$, corresponding to the broken $\PT$-symmetry; (c)-(d) $\sigma=2.5$, corresponding to the unbroken $\PT$-symmetry. Other parameters are $W_0=0.2$ and $\omega=0.5$.} \label{fig3-cos-b}
\end{figure}

For the fixed parameters $W_0=0.2,\,\omega=0.5$, when $\sigma=-310$, for which the $\PT$-symmetry is unbroken, the soliton (\ref{soluc}) is still stable (see Fig.~\ref{fig3-cos-b}(b)), however, when $\sigma=2.5$, in which $\PT$-symmetry is unbroken, the soliton (\ref{soluc}) becomes unstable (see Fig.~\ref{fig3-cos-b}(d)). Thus, for $W_0=0.2,\,\omega=0.5$, the parameter $\sigma$ has significant influence on the linear stability of the soliton solution (\ref{soluc}). A negative $\sigma$ favors to obtain the stable solitons.

\section{Solitons in three-dimensional $\PT$-symmetric potential}

We now consider the 3D NLS equation with the $\PT$-symmetric potential
\bee
i\partial_t \psi\!+\!\nabla^2\psi\!+\![V(x,y,z)\!+\!iW(x,y,z)]\psi\!+\!g|\psi|^2\psi\!=\!0,\,\,\,\,\,
\ene
where $\psi=\psi(x,y,z,t)$ is a complex field with respect to $x,y,z,t\in \mathbb{R}$, $\nabla^2=\partial_x^2+\partial_y^2+\partial_z^2$, $V(x,y,z)$ and $W(x,y,z)$ are both real-valued functions related to external potential and gain-and-loss distribution, respectively, and $g$ is a real constant with $g=\pm 1$. The $\PT$-symmetric potential requires the sufficient (but not necessary) condition $V(x,y,z)=V(-x,-y,-z)$ and $W(x,y,z)=-W(-x,-y,-z)$. The stationary solution can be solved in the form $\psi(x,y,z)=\phi(x,y,z)e^{i\mu t}$, where $\mu$ is the real propagation constant and $\phi(x,y,z)$ satisfies
\bee
\label{ode2}
\mu\phi=\nabla^2\phi+[V(x,y,z)+iW(x,y,z)]\phi+g|\phi|^2\phi.
\ene
\begin{figure}[!t]
	\begin{center}
{\scalebox{0.42}[0.42]{\includegraphics{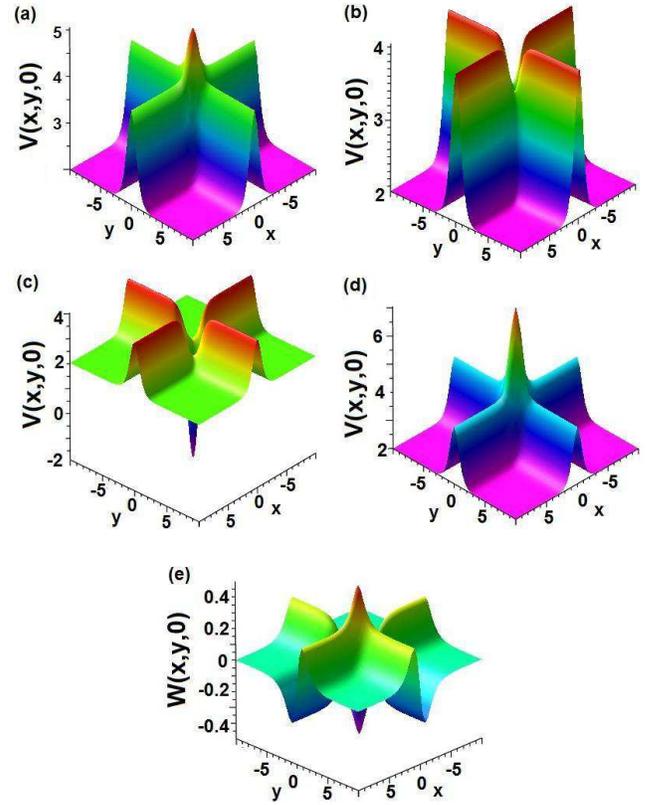}}}
	\end{center}
	\vspace{-0.15in} \caption{\small (color online)  The external potential $V(x,y,0)$ with (a) $g=1$, (b) $g=3$, (c) $g=8$, and (d) $g=-1$; (e) The gain-and-loss distribution $W(x,y,0)$. The parameters are $W_0=0.5$ and $\phi_0=k_x=k_y=k_z=1$.} \label{fig4-vw}
\end{figure}

\begin{figure}[!t]
	\begin{center}
 {\scalebox{0.42}[0.42]{\includegraphics{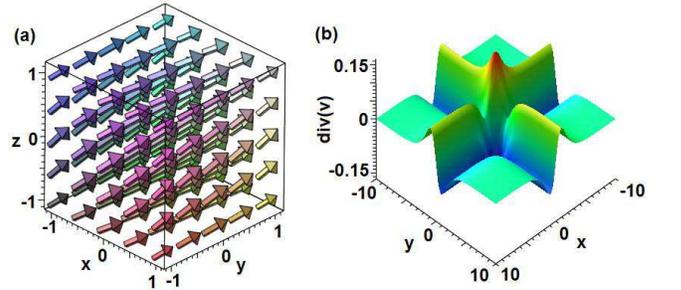}}}
	\end{center}
	\vspace{-0.15in} \caption{\small (color online)  (a) the velocity field $\bv(x,y,z)$ (\ref{vf}); (b) the flux density ${\rm div}\,\bv(x,y,0)$ (\ref{div}). The parameter are $W_0=0.5$ and $k_x=k_y=k_z=1$. } \label{fig6}
\end{figure}

\begin{figure}[!t]
\begin{center}
\vspace{0.15in}	
{\scalebox{0.42}[0.42]{\includegraphics{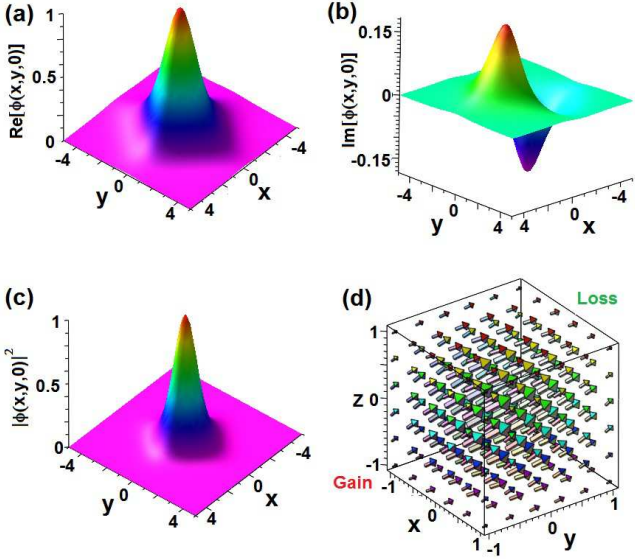}}}
\end{center}
\vspace{-0.15in} \caption{\small (color online) The (a) real part, (b) imaginary part, and (c) the intensity of the soliton solution (\ref{solu2}); (d) The transverse power-flow vector (Poynting vector) $\vec{S}(x,y,z)$. The parameter are $W_0=0.5$ and $k_x=k_y=k_z=1$. } \label{fig5}
\end{figure}

We consider the generalized 3D $\PT$-symmetric Scarff II potential
\bee\label{potend} \begin{array}{l}
 V(x,y,z)=\displaystyle \sum_{\eta=x,y,z}\left(\frac{W_0^2}{9k_\eta^2}+2k_{\eta}^2\right){\rm sech}^2(k_\eta\eta) \vspace{0.1in}\\
 \qquad\qquad\qquad  -g\phi_0^2\prod_{\eta=x,y,z}{\rm sech}^2(k_\eta\eta), \vspace{0.1in}\\
 W(x,y,z)=W_0\sum_{\eta=x,y,z}{\rm sech}(k_\eta\eta)\tanh(k_\eta\eta),
\end{array}\ene
where $k_\eta>0\, (\eta=x,y,z)$ are the wavenumbers in $x,y,z$-directions, respectively, and $W_0,\, \phi_0$ are real constants. In particular, the $\PT$-symmetric potential (\ref{potend}) reduces to the Scarff II potential (\ref{poten}) when $\eta=x$ and $k_x=1$. Since $x,y,z$ are symmetric in $V(x,y,z)$ and $W(x,y,z)$ thus we plot their profiles in $(x,y)$-space with $z=0$ (see Fig.~\ref{fig4-vw}). For the self-focusing nonlinearity $g>0$, the external potential $V(x,y,0)$ exhibits the different profiles as $g$ becomes large (see Fig.~\ref{fig4-vw}(a)-(c)).

For the above-mentioned 3D potential (\ref{potend}) we obtain the exact bright solitons of Eq.~(\ref{ode2})
\bee\label{solu2}
\phi(x,y,z)\!=\!\phi_0{\rm sech}(k_xx){\rm sech}(k_yy){\rm sech}(k_zz)e^{i\varphi(x,y,z)}, \quad
\ene
where $\mu=k_x^2+k_y^2+k_z^2$ and the phase is
 \bee
\varphi(x,y,z)=\frac{W_0}{3}\sum_{\eta=x,y,z}k_\eta^{-2}\arctan[\sinh(k_\eta\eta)].
 \ene
 The real and imaginary parts and intensity of the solution (\ref{solu2}) are shown in Fig.~\ref{fig5}(a)-(c) for $\phi_0=1$ and $W_0=0.5$.

The velocity field $\bv(x,y,z)$ of the solitons (\ref{solu2}) have the form
\bee \label{vf}
\begin{array}{rl}
\bv\!=\!&\!\!\nabla\varphi(x,y,z) \vspace{0.1in}\cr\!=\! &\!\! \frac{W_0}{3}[k_x^{-1}{\rm sech}(k_xx), \, k_y^{-1}{\rm sech}(k_yy),\, k_z^{-1}{\rm sech}(k_zz)],\qquad
\end{array}
\ene
which is shown in Fig.~\ref{fig6}(a) with $W_0=0.5$. It follow from Eq.~(\ref{vf}) that the divergence of velocity field $\bv(x,y,z)$ (alias the flux density) is given by
\bee \label{div}
\begin{array}{rl}
{\rm div}\,\bv(x,y,z)=&\nabla\!\cdot\!\bv(x,y,z)=\nabla^2\varphi(x,y,z)  \vspace{0.1in} \\
=&-\frac{W_0}{3}\sum_{\eta=x,y,z}{\rm sech}(k_\eta\eta)\tanh(k_\eta\eta), \qquad
\end{array}
\ene
which measures the flux per unit area and is dependent on both $W_0$ and space position. Fig.~\ref{fig6}(b) shows the flux density in $(x,y,0)$-space. In addition, we also have the relation
\bee
{\rm div}\,\bv(x,y,z)=\nabla^2\varphi(x,y,z)=-\frac13 W(x,y,z).
\ene

From Eq.~(\ref{div}) we have the following proposition. For the given parameter $W_0>0$, we have:
\begin{itemize}
\item[i)] the fluid flows outward if $f(x,y,z)<0$;

\item[ii)]  the fluid flows inward if $f(x,y,z)>0$;

\item[iii)] the fluid does not flow if $f(x,y,z)=0$.
\end{itemize}
where we have introduced $f(x,y,z)=\sum_{\eta=x,y,z}{\rm sech}(k_\eta\eta)\tanh(k_\eta\eta)$.  For the case $W_0<0$, we also have the corresponding results.

For any complex solution $\phi(x,y,z)=|\phi(x,y,z)|e^{i\varphi(x,y,t)}$, we find that
its corresponding transverse power-flow or Poynting vector and the gradient of phase obey the relation
\bee
 \begin{array}{rl}
 \vec{S}(x,y,z) & = \frac{i}{2}(\phi\nabla\phi^{*}-\phi^{*}\nabla\phi) \vspace{0.1in}\\
& =|\phi(x,y,z)|^2\nabla\varphi(x,y,z).
\end{array}
\ene
The transverse power-flow or Poynting vector related to the solution (\ref{solu2}) is given by
\bee \begin{array}{rl}
\vec{S}(x,y,z)\!\!&=\displaystyle\phi_0^2\prod_{\eta=x,y,z}{\rm sech}^2(k_\eta\eta)\bv(x,y,z) \vspace{0.1in}\\
 &=\dfrac{W_0\phi_0^2}{3}\prod_{\eta=x,y,z}{\rm sech}^2(k_\eta\eta) \vspace{0.1in}\\
 & \times\left[k_x^{-1}{\rm sech}(k_xx), \, k_y^{-1}{\rm sech}(k_yy),\, k_z^{-1}{\rm sech}(k_zz)\right],
\end{array}  \ene
for either $g=1$ or $-1$ (i.e. it does not depend on the sign of the nonlinearity), which is exhibited in Fig.~\ref{fig5}(d). In addition, the conserved power related to the solution (\ref{solu2}) is given by $P(t)=8\phi_0^2/(k_xk_yk_z)$ for either $g=1$ or $g=-1$.\\

\section{Conclusion and discussion}

In conclusion, we have presented a unified theoretical study of the optical bright solitons governed by self-focusing and defocusing NLS equations with $\PT$-symmetric Scarff-like II potentials. Particularly, a $\PT$-symmetric $k$-wavenumber Scarff II potential and a multi-well Scarff II potential are considered, respectively. For the $k$-wavenumber Scarff II potential, the parameter space can be divided into different regions, corresponding to unbroken and broken $\PT$-symmetry and bright solitons with self-focusing and defocusing Kerr nonlinearities. For the multi-well Scarff II potential the bright solitons can be obtained by using a periodically space-modulated Kerr nonlinearity. The linear stability of bright solitons with $\PT$-symmetric $k$-wavenumber and multi-well Scarff II potentials is analyzed in details by using numerical simulations. Stable and unstable bright solitons are found in both regions of unbroken and broken $\PT$-symmetry due to the existence of the nonlinearity. Furthermore, the bright solitons in 3D self-focusing and defocusing NLS equations with a generalized $\PT$-symmetric Scarff II potential are explored. The used method can also be used to study optcial solitons in other $\PT$-symmetric k-wavenumber potentials.

The results we obtained in this work provide a new way of control over soliton stability by using generalized $\PT$-symmetric Scarff II potentials in different parameter domains. This may have potential applications in the field of optical information transmission and processing based on optical solitons in nonlinear dissipative but $\PT$-symmetric systems.

\acknowledgments

\vspace{0.05in}

The authors would like to thank the referees for their invaluable suggestions. This works of ZYY and ZCW were partially supported by NSF-China under Grants No. 61178091, and NKBRPC under Grants No. 2011CB302400. The work of C.H. was supported by NSF-China under Grants No. 11475063.

\end{document}